\documentclass[12pt]{article}

\usepackage{graphicx}
\usepackage{amssymb}
\usepackage{amsthm}
\newtheorem{theorem}{Theorem}

\newtheorem{proposition}[theorem]{Proposition}

\newcommand{\R}{\mathbb{R}}

\title{On the force fields which are homogeneous\\ of degree $-3$}

\author{ Alain Albouy,\\
CNRS, IMCCE, Observatoire de Paris,\\
77, avenue Denfert-Rochereau, 75014 Paris, France\\
  alain.albouy@obspm.fr}

\date{}

\begin{document}

\maketitle

Soon after establishing the famous properties of the $1/r^2$ law of force, Newton described a spiraling orbit of a particle under a central force in $1/r^3$. He also noticed that the addition of a force in $1/r^3$ to another force results in a kind of precession of the orbit (\cite{New}, book 1, proposition 44). In 1842, Jacobi \cite{Jac} gave general results about the force fields which are homogeneous of degree $-3$ and derived from  a potential. More recently, Montgomery \cite{Mon} gave an impressive description of the dynamics of the planar 3-body problem with a force in $1/r^3$. Such homogeneity of the force also appears in Appell's projective dynamics, where  the force is considered together with a constraint (see \cite{Alb}).

Here we deduce a very elementary property: the dynamics defined by a force field which is homogeneous of degree $-3$ can always be reduced, by simply constraining it. This remark is indeed an elegant foundation of Appell's projective dynamics. We will see how it relates to other known properties.

\begin{proposition}\label{p1}

Let $\Omega\subset V$ be an open semi-cone in a finite dimensional real vector space $V$, and $f: \Omega\to V$ be a vector field which is positively homogeneous of degree $-3$. The dynamics of the ordinary differential equation $\ddot q=f(q)$ is reduced by one degree of freedom (i.e.\ by two dimensions) by constraining it to any hypersurface transverse to the rays, the constraint being imposed by means of a central reaction.

\end{proposition}

 Here {\it semi} and {\it positively} refer to the fact that we are only concerned with the {\it half}-lines drawn from the origin of the vector space, that we call the {\it rays}. The term {\it reaction} refers to the familiar mechanical system formed by a particle moving on a surface. In this familiar situation the reaction is normal to the surface. But in our proposition the reaction is {\it central}, i.e.\ ``radial'', i.e.\ carried by the ray. The existence and uniqueness theorems for the solution of such kind of constrained system are easy. Their proofs do not depend on the particular choice concerning the direction of the reaction, provided that this direction is fixed in advance and transverse to the hypersurface.

\begin{proof}

We write the equation of the hypersurface $h(q)=1$, where $h:\Omega\to\, ]0,+\infty[$ is a positively homogeneous function of degree 1. We denote by $q_1=q/h(q)$ the central projection of $q$ on the hypersurface. We will show that $q_1$ follows some trajectory of the system defined by the constraint and by the force field $f$.

We start with the given equation $\ddot q=f(q)$. We compute $\dot q_1=h^{-2}(h\dot q-\dot h q)$. Instead of differentiating again with respect to the time $t$, we introduce a change of time depending only on the position $q$. The corresponding differentiation on any quantity $r$ is denoted by $r'$ and the change of time is defined by the formula $r'=h^2\dot r$. We get $q_1'=h\dot q-\dot h q$, $\dot {q_1'}=h\ddot q-\ddot h q$ and $q_1''=h^3\ddot q-h^2\ddot h q$. But $h^3\ddot q=h^3f(q)=f(q_1)$ according to the degree of homogeneity of the force field $f$.
The final equation is $q_1''=f(q_1)+\lambda q_1$, where $\lambda=-h^3\ddot h$. The value of $\lambda$ should be rather thought of as determined by the constraint: $q_1$ remains on the hypersurface, which determines uniquely the value of the multiplier $\lambda$.

\end{proof}

This reduction process is not standard. The reduction by two dimensions does not involve a constant of motion. We can describe it as the effect of two vector fields $Y$ and $Z$ related with the vector field $X$ defined by our ordinary differential equation. The three vector fields are characterised by $\partial_X q=\dot q=p$, $\partial_X p=\dot p=f(q)$,  $\partial_Yq=q$, $\partial_Yp=-p$, $\partial_Zq=0$, $\partial_Zp=q$. The Lie brackets $[X,Y]=2X$, $[Y,Z]=2Z$, $[Z,X]=Y$  show that the subspaces generated at each $(q,p)$ by $X$, $Y$ and $Z$ form an integrable distribution in the sense of the Frobenius [-Stefan-Sussmann] theorem. Note also that these brackets define a Lie algebra $sl_2$. The 3-dimensional integral manifolds intersect our constraint along curves, which are the trajectories of the constrained system.

A force field $f$ with degree of homogeneity  $\alpha$ defines an $X$ which satisfies the commutation relation $[X,Y_\beta]=(1-\beta)X$, where $Y_\beta$ is defined by $\partial_{Y_\beta}q=q$, $\partial_{Y_\beta}p=\beta p$, and where $2\beta=\alpha+1$. If $\alpha\neq -3$, nothing replaces the vector field $Z$, and we can only reduce by one dimension.

Proposition \ref{p1} is the fastest way to introduce projective dynamics. If we start with a dynamics defined by a force field on an affine space of dimension $n$, we can embed this space as an affine hyperplane in a vector space $V$ of dimension $n+1$, and  extend the force to $V$ by homogeneity of degree $-3$. Then we constrain this homogeneous force field  to another hypersurface (another ``screen''), thus producing another system which is very simply related with the initial one. Many dynamical properties are thus preserved by central projection. We already know that many geometrical properties are preserved by central projection, and this remark is the foundation of projective geometry. Thus, we should similarly consider that there is a projective dynamics, which extends projective geometry to the motions generated by force fields.

Applying this construction to the two fixed centres problem allows deducing the well-known integrability of this problem from purely geometric considerations. By choosing a convenient quadric as the other screen the question reduces to considerations on the intersections of a plane with two cylinders (see \cite{Alb}).

Facts related to Proposition \ref{p1} are known in the case where $f$ is derived from a potential (see \cite{AlC}, pp.\ 161, 169, 172  and \cite{BKM}). The following observation may be new.

\begin{proposition}\label{p2}

If the vector space $V$ is endowed with an inner product, if the force field $f$ of Proposition \ref{p1} is the gradient of a function $U:\Omega\to \R$ with respect to this inner product, and if we constrain $f$,   by means of a central reaction, to the intersection $S$ of $\Omega$ with the unit sphere, then the multiplier $\lambda$ associated to the constraint is the energy multiplied by $-2$.

\end{proposition}

Here the central reaction is  normal.  We have a natural constrained system on $S$. The potential is the restriction of $U$ to $S$. If we start with such a natural system on $S$, we form the unique extension of the potential in a function $U$ on $\Omega$ which is positively homogeneous of degree $-2$.  Note that $f=\nabla U$ will not be tangent to the sphere, which does not affect the dynamics but does affect the value of the multiplier $\lambda$.

\begin{proof}

To determine $\lambda$ in the equation $\ddot q=\nabla U+\lambda q$ we differentiate twice the constraint $\langle q,q\rangle=1$. We get $0=\langle q,\dot q\rangle$ and $0=\langle q,\nabla U+\lambda q\rangle+\langle \dot q,\dot q\rangle=-2U+\lambda+\langle\dot q,\dot q\rangle$.

\end{proof}

Proposition \ref{p2} plays an interesting role in the relation discovered by Kn\"orrer between the Neumann potential on the sphere and the geodesics on an ellipsoid (see \cite{Kno}, \cite{BoM}). We will exhibit an intermediate problem which clarifies this relation as well as the integrability of the Neumann potential.  Consider a symmetric positive definite $G: V\to V^*$ and the vector field on $\Omega=V\setminus\{0\}$
$$f(q)=\frac{Mq}{\langle Gq,q\rangle^2}, \qquad\hbox{where } M:V\to V\hbox{ is such that }GM={}^tM G.$$
In words, the linear map $M$ is symmetric with respect to the inner product $G$. We may make explicit this symmetry in two ways: through a symmetric $A:V\to V^*$ such that $M=G^{-1}A$ or through a symmetric $B:V\to V^*$ such that $M=B^{-1}G$. For simplicity of exposition we assume that $A$ is positive definite (and then so is $B=GA^{-1}G$).

The formula $M=G^{-1}A$ suggests to endow $V$ with the inner product $G$ and to observe that $f$ is, up to a central force, the gradient of the function $\langle Gq,q\rangle^{-2}\langle Aq,q\rangle/2$. By constraining the dynamics to the sphere $\langle Gq,q\rangle=1$, we get the Neumann potential.

The formula $M=B^{-1}G$ suggests to endow $V$ with the inner product $B$ and to observe that $f$ is the gradient of the function $-\langle Gq,q\rangle^{-1}/2$. Constraining the dynamics to $\langle Bq,q\rangle=1$, we get our intermediate problem, whose integrability was established by Braden (see \cite{Bra}, \cite{Woj}). This is a natural system on the sphere, defined by a potential which is the inverse of a quadratic form (while the Neumann potential is a quadratic form).

\medskip
\centerline{\includegraphics [width=120mm]
{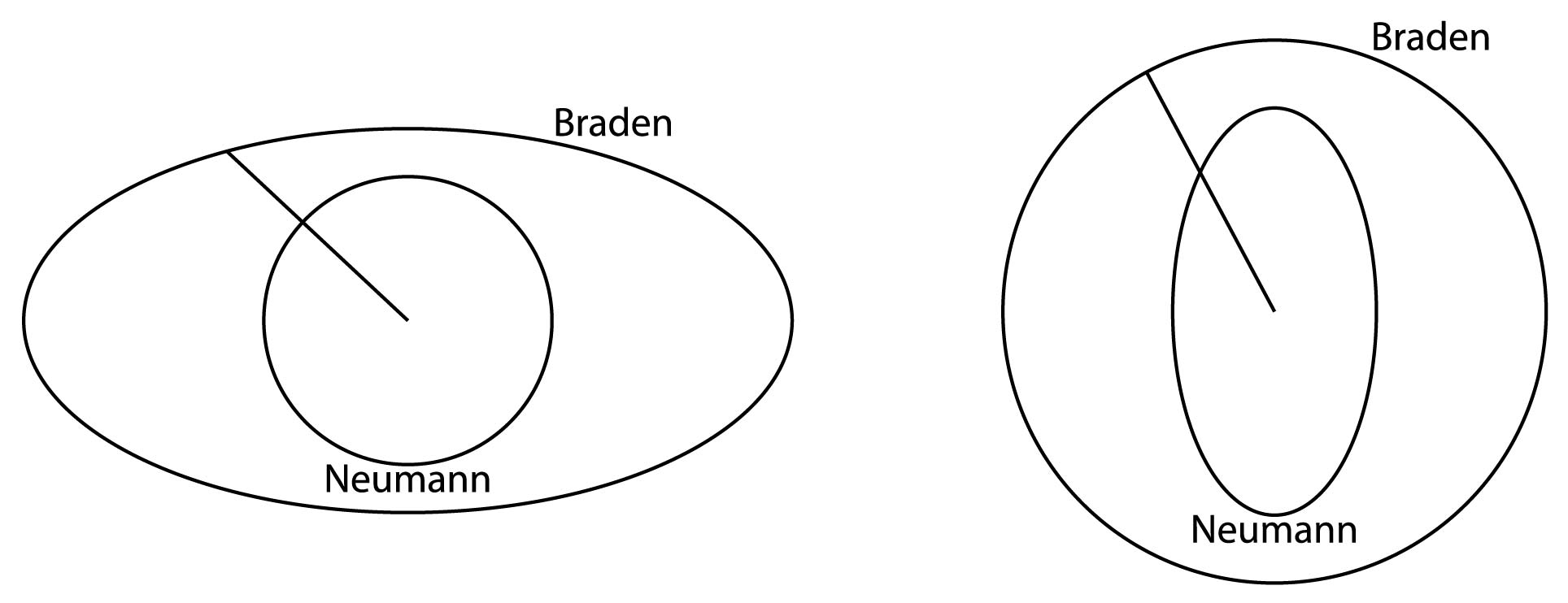}}
\medskip

We can deduce the integrability of both the Neumann potential and our intermediate problem from their correspondence through central projection and change of time. They are {\it quasi-bi-Hamiltonian} systems, as already claimed about the Neumann potential in \cite{Bla}, \cite{Ped}, \cite{Al1}. This last reference also connects this remark to the works \cite{Lu1}, \cite{MaT}, \cite{Ta1}.

Let us consider the Jacobi problem on the ellipsoid. The motion of a particle $Q$ on the ellipsoid $\langle AQ,Q\rangle=1$ embedded in the Euclidean vector space $(V,G)$, under the potential $\nu\langle GQ,Q\rangle/2$, is defined by the equation
$$\ddot Q=\mu MQ+\nu Q.$$
Here $\mu$ is a multiplier. The case $\nu=0$ defines the geodesic motion on the ellipsoid. The addition of this potential was already considered by Jacobi, and again by Moser \cite{Mos} in connection with Kn\"orrer's work. Differentiating the constraint three times, we find Joachimsthal's constant of motion in the form $\eta=\mu\langle AQ,MQ\rangle^2$. The motion of $q=MQ$ is constrained by $\langle Bq,q\rangle=1$ and satisfies the equation
$$\ddot q=\frac{\eta}{\langle Gq,q\rangle^2} Mq+\nu q.$$
This constraint and this equation also define our intermediate problem. But $\eta$ and $\nu$ have a different interpretation in both problems. In the Jacobi problem, $\nu$ is a parameter and $\eta$ is a multiplier which appears to be a constant of motion. In our intermediate problem $\eta=1$ and $\nu$ is a multiplier which, according to Proposition \ref{p2}, is a constant of motion. Any orbit of a problem is an orbit of the other problem for some choice of a parameter.

This is similar to what was explained by Kn\"orrer and Moser, except that they needed a change of the time parameter and we did not. The motion on a sphere under the inverse of a quadratic potential is thus closer to the Jacobi problem than the motion on the sphere under a quadratic potential.

The introduction of our intermediate problem allows decomposing the Gauss map $Q\mapsto MQ/\|MQ\|$ introduced by Kn\"orrer, into two steps: $Q\mapsto q\mapsto q/\|q\|$. Kn\"orrer's change of time appears in the second step as associated to the central projection from our intermediate problem to the Neumann problem. It satisfies the rule, discovered by Appell, which associates a change of time to a central projection.

{\bf Acknowledgements.} This work benefited from discussions with Alexey Borisov, Alain Chenciner, Yuri Fedorov, Bo\v zidar Jovanovi\'c, Hans Lundmark and Ivan Mamaev.


\begin{thebibliography}{99}

\parskip0pt

\itemsep0pt

\bibitem{Alb} A. Albouy, There is a projective dynamics, {\it EMS Newsletter}, 89 (2013), pp.\ 37--43

\bibitem{Al1} A. Albouy, Projective dynamics of a classical particle or multiparticle system, {\it Oberwolfach reports}, 4-3 (2007), pp.\ 1926--1928

\bibitem{AlC} A. Albouy, A. Chenciner,  Le probl\`eme des N corps et les
distances mutuelles, {\it Inventiones Mathematicae}, 131 (1998),
pp.\ 151--184

\bibitem{Bla} M. B{\l}aszak,  Bi-Hamiltonian separable chains on Riemannian manifolds, {\it Physics Letters A}, 243 (1998), pp.\ 25--32

\bibitem{Bra} H.W. Braden, A completely integrable mechanical system, {\it Letters in Mathematical Physics}, 6 (1982), pp.\ 449--452

\bibitem{BKM}  A.V. Borisov, A.A. Kilin, I.S. Mamaev, Multiparticle systems. The algebra of integrals and integrable cases, {\it Regular and Chaotic Dynamics}, 14 (2009), pp.\ 18--41

\bibitem{BoM}  A.V. Borisov, I.S. Mamaev, Isomorphisms of geodesic flows on quadrics, {\it Regular and Chaotic Dynamics}, 14 (2009), pp.\ 455--465

\bibitem{Jac} C.G.J. Jacobi, {\it Lectures on Dynamics}, transl.\ K.\ Balagangadharan, New Delhi, Hindustan Book Agency, 2009, lecture 4

\bibitem{Kno} H.\ Kn\"orrer, Geodesics on quadrics and a mechanical problem of C.\ Neumann, {\it J.\ reine angew.\ Math., }334 (1982), pp.\ 69--78

\bibitem{Lu1} H. Lundmark, Higher-dimensional integrable Newton systems with quadratic integrals of motion, {\it Studies in Applied Math., }110 (2003), pp.\ 257--296

\bibitem{MaT} V.S. Matveev, P.J. Topalov, Trajectory equivalence and corresponding integrals, {\it Regular and Chaotic Dynamics}, 3 (1998), pp.\ 30--45

\bibitem{Mon} R. Montgomery,  Fitting hyperbolic pants to a three-body
problem, {\it Ergod.\ Th.\ \& Dynam.\ Sys., }25 (2005), pp.\ 921--947

\bibitem{Mos} J. Moser, Integrable Hamiltonian systems and spectral theory, {\it Fermi Lectures}, Accademia nazionale dei Lincei, Scuola normale superiore, Pisa, 1981

\bibitem{New} Isaac Newton, {\it The {\rm Principia}. A New Translation {\rm preceded by} A guide to Newton's {\rm Principia}, by I.\ Bernard Cohen}, Berkeley, University of California Press, 1999

\bibitem {Ped} M. Pedroni, Bi-Hamiltonian aspects of the separability of the Neumann system, {\it Theoretical and Mathematical Physics}, 133 (2002), pp.\ 1722--1727

\bibitem{Ta1} S. Tabachnikov, Ellipsoids, complete integrability and hyperbolic geometry, {\it Moscow Math.\ J., }2 (2002), pp.\ 185--198

\bibitem{Woj} S. Wojciechowski, Integrable one-particle potentials related to the Neumann system and the Jacobi problem of geodesic motion on an ellipsoid, {\it Physics Letters A}, 107 (1985), pp.\ 106--111







\end{thebibliography}
\end{document}